\documentclass[11pt,twoside]{article}
\usepackage[pdftex]{graphicx}
\usepackage{amsmath}
\usepackage{amssymb}
\usepackage{array,dcolumn}
\usepackage{textgreek}
\usepackage[utf8]{inputenc}
\usepackage[T1]{fontenc}
\usepackage{mathptmx}
\usepackage{amsfonts,wrapfig,booktabs,makecell}
\usepackage{hyperref,url}
\usepackage{csquotes}
\begin{document}
\newcommand{\pst}{\hspace*{1.5em}}

\newcommand{\rigmark}{\em Journal of Russian Laser Research}
\newcommand{\lemark}{\em Volume XX, Number X, 20XX}

\newcommand{\be}{\begin{equation}}
\newcommand{\ee}{\end{equation}}
\newcommand{\bm}{\boldmath}
\newcommand{\ds}{\displaystyle}
\newcommand{\bea}{\begin{eqnarray}}
\newcommand{\eea}{\end{eqnarray}}
\newcommand{\ba}{\begin{array}}
\newcommand{\ea}{\end{array}}
\newcommand{\arcsinh}{\mathop{\rm arcsinh}\nolimits}
\newcommand{\arctanh}{\mathop{\rm arctanh}\nolimits}
\newcommand{\bc}{\begin{center}}
\newcommand{\ec}{\end{center}}

\thispagestyle{plain}

\label{sh}


\begin{center} {\Large \bf
\begin{tabular}{c}
ON THE HELICITY OF RADIATIVE
\\[-1mm]
ELECTROMAGNETIC FIELDS
\end{tabular}
 } \end{center}

\bigskip

\bigskip

\begin{center} {\bf
R.M. Feshchenko$^{1}$, and A.V. Vinogradov$^{1}$
}\end{center}

\medskip

\begin{center}
{\it
$^1$P.N. Lebedev Physical Institute, Russian Academy of Sciences, 53 Leninski Prospekt\\
Moscow, 119991, Russia
}
\smallskip

$^*$Corresponding author e-mail: rusl@sci.lebedev.ru\\
\end{center}

\begin{abstract}\noindent
A relativistically invariant expression for the magnetic and electric helicities of free electromagnetic field through the currents, which created those field, are derived. It is demonstrated that for radiative fields (e.g. laser pulses) the both helicities are conserved and equal to each other. Moreover the conservation of the helicity means that in radiative electromagnetic fields the electric and magnetic vectors are always orthogonal to each other. Proceeding to the quantum theory, it is shown that the corresponding quantum helicity field operator is equal to the sum of projections of spins of all photons onto their momentums. Finally, as an example, the expressions for the total helicity, number of photons and the total spin momentum of the field of a dipole electromagnetic pulse are obtained and discussed.
\end{abstract}

\medskip


\section{Introduction}
\pst
As it is well known that for any free electromagnetic field a quantity can be introduced called helicity, which serves to characterize the field's handiness \cite{afanasiev1996helicity}. The helicity of the freely propagating electromagnetic pulses, including femtosecond laser pulses, can be important parameter when one needs to produced polarized electron \cite{PhysRevLett.128.174801} or positron beams \cite{PhysRevLett.125.044802}. Moreover, laser beams with non-zero helicity can be used in all-optical helicity-dependent switching of magnetic materials \cite{wang2018all}.

It should be noted that helicity can be defined through the magnetic field and conventional vector potential -- so called magnetic helicity or through the electric field and corresponding magnetic vector potential -- so called electrical helicity \cite{trueba1996electromagnetic}. Thus the definition of helicity is generally is not gauge invariant. Even more, the mere conservation of the helicity depends on its definition -- it can be a conserved quantity only of a radiative field\footnote{Radiative field is a free field emitted by a temporally and spatially finite system of currents.} or of an arbitrary free field. The helicity is also related to the spin angular moment of the free electromagnetic field, which serves as the Noether current for it, the helicity being the corresponding Noether charge \cite{cameron2017chirality,anderson1987angular}. 

The definition of helicity can be further generalized to electromagnetic fields in the medium although its definition can differ from that in free space \cite{guasti2019chirality}. The helicity of the quantized electromagnetic field and its relation to the Stokes parameters were considered in \cite{bialynicki2019helicity}.

In our previous works \cite{feshchenko2019number,FeshchenkoVinogradov2018} we derived relativistically invariant expressions for the total number of photons \cite{Zeldovich1966} and total spin momentum \cite{cameron2017chirality} of a classical radiative electromagnetic field through the currents, which created this field. We also explored how those expressions are related to each other and to the quantum field operators of number of photons and total spin angular momentum of the electromagnetic field. In those works we, however, chose not to consider the helicity and, therefore, the question of how it can be expressed through the currents was left unanswered. 

In this article we fill in this gap by providing general expressions of differently defined electromagnetic helicities through the currents, which emitted those fields, and demonstrate that for radiative fields the helicity is a conserved quantity regardless of its definition. We also explore the question of how the quantum operator of helicity is expressed through the creation and annihilation operators of the electromagnetic field much like we did for the total number of photons and total spin momentum in \cite{feshchenko2019number}.  

\section{Helicity radiated by a finite system of currents}
\pst
As it was mentioned above the helicity of the electromagnetic field does not have singular meaning and is generally gauge dependent. The most frequently encountered definition involves the electromagnetic field expressed through its conventional 4D vector potential $A^i$ -- so called magnetic helicity. There also exists the reciprocally defined electric helicity, which we will discuss below.

At the first step let us derive the expression for the total magnetic helicity $h_M$ of free electromagnetic field, which is defined as the flux of the helicity 4D vector 
\begin{equation}
\label{1a}
h^i=\frac{1}{8\pi}\varepsilon^{ikmp}A_{k}F_{mp}
\end{equation}
through a hyperplane $\cal S$ in the Minkowski space defined by a unity 4D vector $n^i$ orthogonal to it (see also \cite{zangwill2013modern} p. 327)
\begin{equation}
\label{1b}
h_M=\frac{1}{8\pi}\int\varepsilon^{ikmp}A_{k}F_{mp}dS_i=\frac{1}{4\pi}\int(\mathbf{AH})d^3V,
\end{equation}
where $n^in_i=1$ (summation is implied for the repeated upper and lower indexes), $A^i=(A^0,\mathbf{A})$ is the vector potential of the electromagnetic field emitted by the 4D current $j^l$, $F^{ik}$ is the electromagnetic field tensor, $\mathbf{H}$ is the 3D magnetic field vector, $\varepsilon^{ikmp}$ is 4D Levi--Civita antisymmetric tensor and $g^{ik}=\mathrm{diag}(1,-1,-1,-1)$ is the Minkowski space metric tensor. Everywhere it is also assumed that speed of light $c=1$. 

The last part of (\ref{1b}) was obtained assuming that the vector $n^i=(1,0,0,0)$ and substituting integration over hyperplane $\cal S$ with integration over the 3D space. The definition (\ref{1a}) is different from the conventional one by a factor $1/(4\pi)$ \cite{cameron2017chirality}. Such a choice is explained by an observation that the spatial components of $h^i=(h^0,\mathbf{s})$ defined by (\ref{1a}) coincide with the spin momentum density of the electromagnetic field \cite{cameron2017chirality}.

Since the 4D divergence of the vector (\ref{1a}) is in a general case not zero and equal to the second invariant of the electromagnetic field
\begin{equation}
\label{1b1}
\frac{\partial h^i}{\partial x^i}=\frac{\varepsilon^{ikmp}F_{ik}F_{mp}}{16\pi}=\frac{\mathbf{EH}}{4\pi},
\end{equation}, 
where $\mathbf{E}$ is the 3D electrical field vector, the value of $h_M$ generally depends on the choice of hyperplane $\cal S$ even for a free electromagnetic field. It though does not depend on the hyperplane choice in one important case of the radiative electromagnetic field. To calculate the helicity (\ref{1b}) in this case let us note that the 4D Fourier transform of the 4D potential $A^l(k)$ for the radiative electromagnetic field can be expressed through the 4D Fourier transform of the 4D current $j^l(k)$ using the electromagnetic Pauli--Jordan function \cite{Bogolubov1982} as was done in \cite{feshchenko2019number} 
\begin{equation}
\label{1c}
A^i(k)=-i8\pi^2\mathrm{sign}(k^0)\delta(k^2)j^i(k),
\end{equation}
where the 4D current $j^i=(j^0,\mathbf{j})$ is considered to exist only in a finite spatio-temporal domain, the potential $A^i$ is taken in the Lorentz gauge and the 4D Fourier transforms in (\ref{1c}) are defined as
\begin{eqnarray}
A^i(k)&=&\int A^i(x)e^{ik_mx^m}\,d^4x,\label{1d1}\\
j^i(k)&=&\int j^i(x)e^{ik_mx^m}\,d^4x,\label{1d2}
\end{eqnarray}
where $x^i=(t=x^0, \mathbf{r})$ are 4D coordinates, $j^0$ is the volume charge density and $\mathbf{j}$ is the 3D current vector, $k^i=(k^0,\mathbf{k})$ is the 4D wave vector ($k^2=k^\nu k_\nu$, $k^0=|\mathbf{k}|=\omega$ is the frequency), $\mathrm{sign}(x)=2\theta(x)-1$ is the sign function and $\theta(x)$ is the $\theta$-function (unity step function). 

Since in the formula (\ref{1c}) the positive and negative frequency parts have the equal absolute values, it is sufficient to consider only the positive frequency part 
\begin{equation}
\label{1e}
A^i(k)=-i8\pi^2\theta(k^0)\delta(k^2)j^i(k).
\end{equation}
The contribution of the negative frequency part to the total helicity will be equal to that of the positive frequency part. 

The total helicity the electromagnetic field can then be found by integrating in expression (\ref{1a}) over a hyperplane $\cal S$, while taking into account that the coordinate dependent potential $A^i(x)$ is defined as the inverse Fourier transform of $A^i(k)$
\begin{equation}
\label{1f}
A^i(x)=\frac{1}{(2\pi)^4}\int A^i(k)e^{-ik_mx^m}d^4k.
\end{equation}
The electromagnetic field tensor $F^{ik}$ can then be written as \cite{landau2013classical} 
\begin{equation}
\label{1g}
F^{ik}=\frac{\partial A^k}{\partial x_i}-\frac{\partial A^i}{\partial x_k}.
\end{equation}
Let us now assume that the hyperplane ${\cal S}$ lies outside the domain where the currents are present. In other words, we consider only free electromagnetic fields produced by some currents in the past, which have disappeared since then. The hyperplane ${\cal S}$ can be defined by the following equation
\begin{equation}
\label{1g1}
n_m(x^m-y^m)=0,
\end{equation}
where $n^i$ is a unity 4D vector orthogonal to the hyperplane and $y^i$ is an arbitrary 4D vector. Then the integral expression (\ref{1b}) can be rewritten as
\begin{equation}
h_M=\frac{1}{8\pi}\int\varepsilon^{ikmp}A_{k}F_{mp}n_i\delta((x^m-y^m)n_m)d^4x=\frac{1}{16\pi^2}\int\limits_{-\infty}^{+\infty}d\zeta\int\varepsilon^{ikmp}A_{k}F_{mp}\exp(i\zeta (x^m-y^m)n_m)d^4x.\label{1g2}
\end{equation}

After substituting (\ref{1e}) into expressions (\ref{1f}), (\ref{1g}) and (\ref{1g2}), then multiplying (\ref{1g2}) by factor $2$ to account for the contribution of the omitted negative frequencies and integrating by $d^4x$ the formula (\ref{1g2}) turns into the following expression
\begin{equation}
h_M=\frac{i}{\pi^2}\int d^4k\int d\zeta\int d^4k'\delta(k-k'-\zeta n)\theta(k^0)\theta(k'^0)\delta(k^2)\delta(k'^2)e^{-iy^mn_m\zeta}\varepsilon^{ikmp}n_ik_kj_m(k)j^*_p(k').\label{1g3}
\end{equation}
Finally, after integrating by $d\zeta$ and $d^4k'$ in (\ref{1g3}), the total helicity of the radiative field produced by a finite system of currents can be found using the following formula 
\begin{equation}
\label{1h}
h_M=\frac{i}{2\pi^2}\int\theta(k^0)\frac{\delta(k^2)}{nk}\varepsilon^{ikmp}n_i k_k j_m(k)j^*_p(k)\,d^4k, 
\end{equation}
which expresses it through the 4D Fourier harmonics of the 4D current with positive frequencies.

Formula (\ref{1h}) at the first glance depends on the 4D vector $n^i$. However we can use a gauge transformation of potentials $A^i\to A^i+c(k^i)k^i$ to eliminate simultaneously the temporal and longitudinal components of the vector potential $A^i$ and to express the magnetic helicity through only the transversal components of the current $\mathbf{j}_\perp$ as
\begin{equation}
\label{1j}
h_M=\frac{i}{4\pi^2}\int\frac{\mathbf{k}}{\left(k^0\right)^2}(\mathbf{j}_\perp(\mathbf{k})\times\mathbf{j}_\perp^*(\mathbf{k}))\,d^3k,
\end{equation}
which does not depend on $n^i$ and is consequently a conserved quantity. This also means that the divergence (\ref{1b1}) is zero and therefore for radiative fields the electrical vector is always orthogonal to the magnetic vector.

Let us now calculate the electrical helicity, which can be defined by subjecting the field to the formal transformation
\begin{equation}
\mathbf{E}\to\mathbf{H},\quad \mathbf{H}\to-\mathbf{E},
\label{1k}
\end{equation}
which preserves the Maxwell equations. This transformation also turns the 4D vector potential $A^i$ into the so called magnetic vector potential $\tilde A^i$, which depends on the magnetic current $\tilde j^i$ as in (\ref{1e}). In the 4D form transformation (\ref{1k}) replaces the field tensor $F^{ik}$ with its dually conjugate tensor 
\begin{equation}
F^{*ik}=\frac{1}{2}\varepsilon^{ikmp}{F_{mp}}.
\label{1l}
\end{equation}
Since both $\tilde A^i$ and $A^i$ are potentials of the same field and taking into account (\ref{1e}) and (\ref{1l}) one can obtain the following relation between currents $\tilde j^i$ and $j^i$
\begin{equation}
k^i\tilde{j}^k-k^k\tilde{j}^i=\varepsilon_{ikmp}{k^mj^p},
\label{1l1}
\end{equation}
which means that in case of radiative field the transversal components of magnetic current $\tilde{\mathbf{j}}_{\perp}$ are rotated by $\pi/2$ relative to $\mathbf{j}_{\perp}$ in the plane perpendicular to $\mathbf{k}$. This directly follows from the equalities $k^2=0$ and $kj=k\tilde{j}=0$ and means that only the transversal components of both currents are well defined, while the longitudinal and temporal components can be chosen arbitrary.

The electrical helicity $h_E$ is then defined by an expression similar to (\ref{1b})
\begin{equation}
h_E=\int\tilde h^{i}\,dS_i=\frac{1}{8\pi}\int\varepsilon^{ikmp}\tilde A_{k}F^{*}_{mp}\,dS_i=-\frac{1}{4\pi}\int(\tilde{\mathbf{A}}\mathbf{E})d^3V.\label{1m}
\end{equation}
The divergence of the corresponding electrical helicity 4D vector $\tilde h^i$ is equal to the second invariant taken with the opposite sign
\begin{equation}
\label{1m1}
\frac{\partial \tilde h^i}{\partial x^i}=\frac{\varepsilon^{ikmp}F^{*}_{ik}F^{*}_{mp}}{16\pi}=-\frac{\mathbf{EH}}{4\pi},
\end{equation} 
which should be compared to (\ref{1b1}). So, for an arbitrary free electromagnetic field the divergence of the sum of the magnetic and electrical helicities $h^i+\tilde h^i$ will be zero and their sum $h_M+h_E$ is therefore conserved.

As to radiative fields one can prove (by repeating the calculation of (\ref{1g2})--(\ref{1j}) and using the definition (\ref{1m})) that $h_E$ can be expressed by the formula (\ref{1j}), where the components $j_{\perp}^i$ are substituted with $\tilde{j}_{\perp}^i$. Thus the total electrical helicity is also an invariant quantity much like the total magnetic helicity and both are conserved separately. Moreover, as it was mentioned above $\tilde{\mathbf{j}}_{\perp}$ is rotated by $\pi/2$ with respect to $\mathbf{j}_{\perp}$, which means that $h_E=h_M$ and therefore any radiative field has in fact only one helicity $h=h_E=h_M$.

\section{Helicity operator and connection with the spin momentum}
\pst
To make transition to the quantum electrodynamics let us introduce the following two functions of the 3D wave vector $\mathbf{k}$
\begin{eqnarray}
a^-_s(\mathbf{k})&=&\frac{1}{2\pi\sqrt{k^0\hbar}}j_{\perp,s}(k^0,\mathbf{k}),\label{2f1}\\
a^+_s(\mathbf{k})&=&\frac{1}{2\pi\sqrt{k^0\hbar}}j^*_{\perp,s}(k^0,\mathbf{k}),\label{2f2}
\end{eqnarray}
where index $s=1,2$ refers to the two non-zero orthogonal components of $\mathbf{j}_\perp$. Now the total helicity (\ref{1j}) of a radiative field can be expressed as 
\begin{equation}
h=h_M=i\hbar\int\left[a^+_2(\mathbf{k})a^-_1(\mathbf{k})-a^+_1(\mathbf{k})a^-_2(\mathbf{k})\right]\,d^3k.\label{2g1}
\end{equation}
As was shown in \cite{feshchenko2019number} in formula (\ref{2g1}) the functions $a^-_s$ and $a^+_s$ should be interpreted as annihilation and creation operators of a photon with momentum $\mathbf{k}$ and linear polarization $s$, which is signified by adding a hat to them. These operators must obey a Bose type commutative relation
\begin{equation}
\label{2h}
\{\hat{a}^-_s(\mathbf{k}),\hat{a}^+_{s'}(\mathbf{k}')\}=\delta_{ss'}\delta(\mathbf{k}-\mathbf{k}').
\end{equation}
Substituting the functions $a^-_s$ and $a^+_s$ with the operators $\hat{a}^-_s$ and $\hat{a}^+_s$ in (\ref{2g1}) one gets the operator $\hat{h}$ of the helicity of electromagnetic field.

Since expression (\ref{2g1}) is diagonal, it can be diagonalized by an unitary transformation (see \cite{feshchenko2019number})
\begin{eqnarray}
a^-_1(\mathbf{k})&=&\frac{1}{\sqrt{2}}(b^-_1(\mathbf{k})+b^-_2(\mathbf{k})),\label{2i1}\\
a^+_2(\mathbf{k})&=&-\frac{i}{\sqrt{2}}(b^+_1(\mathbf{k})-b^+_2(\mathbf{k})),\label{2i2}\\
a^-_2(\mathbf{k})&=&\frac{i}{\sqrt{2}}(b^-_1(\mathbf{k})-b^-_2(\mathbf{k})),\label{2i3}\\
a^+_1(\mathbf{k})&=&\frac{1}{\sqrt{2}}(b^+_1(\mathbf{k})+b^+_2(\mathbf{k})),\label{2i4}
\end{eqnarray}
where $b^-_s$ and $b^+_s$ should be similarly interpreted as the annihilation and creation operators of a photon with momentum $\mathbf{k}$ and a right-handed ($s=1$) or left-handed ($s=2$) circular polarization with the helicity being equal to $1$ or $-1$, respectively. The operators $b^-_s$ and $b^+_s$ obey the same commutative relation (\ref{2h}). The helicity operator $\hat{h}_M$ can finally be expressed as 
\begin{equation}
\hat{h}_M=\hbar\int\left[\hat{b}^+_1(\mathbf{k})\hat{b}^-_1(\mathbf{k})-\hat{b}^+_2(\mathbf{k})\hat{b}^-_2(\mathbf{k})\right]\,d^3k,
\label{2j1}
\end{equation}
being diagonal now. The expressions similar to (\ref{2j1}) can also be found in \cite{bialynicki2019helicity,Bogolubov1982}.

Comparing (\ref{2j1}) with the corresponding expressions for quantum operators of the total number of photons and total spin momentum $\hat{\mathbf{S}}$ from \cite{feshchenko2019number} 
\begin{eqnarray}
\hat{N}&=&\int\left[\hat{b}^+_1(\mathbf{k})\hat{b}^-_1(\mathbf{k})+\hat{b}^+_2(\mathbf{k})\hat{b}^-_2(\mathbf{k})\right]\,d^3k,\label{2k1}\\
\hat{\mathbf{S}}&=&\hbar\int\mathbf{n}\left[\hat{b}^+_1(\mathbf{k})\hat{b}^-_1(\mathbf{k})-\hat{b}^+_2(\mathbf{k})\hat{b}^-_2(\mathbf{k})\right]\,d^3k,\label{2k2}
\end{eqnarray}
one can see that the helicity is indeed an algebraic sum of the projections of spins of photons onto their momentums or using an alternative interpretation -- a difference between the number of right- and left-handed photons \cite{trueba1996electromagnetic}. 

From (\ref{2j1}) and (\ref{2k1})--(\ref{2k2}) it is clear that there exists a direct correspondence between the classical expression for the total helicity and respective quantum operator, similar to the correspondence that exists between the total number of photon, total spin and their operators. So, it is clear that there is an intrinsic connection between the spin momentum and helicity of the radiative electromagnetic field because the former can be interpreted as the sum of projections of spins of photons present in the radiative field onto their momentums.

\section{Helicity of a dipole pulse}
\pst
Let us consider a toy model of radiative field corresponding to a magnetic dipole pulse. Such a choice is explained by an observation \cite{gonoskov2012dipole} that dipole pulses archive the highest field strength in their centers of all conceivable finite pulses. A magnetic dipole pulse can be defined using the magnetic Hertz vector \cite{artyukov2020collapsing}
\begin{equation}
\label{3a}
\mathbf{M}(t,\mathbf{r})=\mathbf{M}_0\frac{f(t+r)-f(t-r)}{r},
\end{equation}
where $=\mathbf{M}_0$ is an arbitrary constant vector, $f(s)$ is an arbitrary smooth function and $r=|\mathbf{r}|$. The vector potential is $\mathbf{A}=\nabla\times\mathbf{M}$ and the scalar potential is zero. For the electric dipole pulse the helicity will be exactly the same as for the magnetic dipole owning to the transformation (\ref{1k}) and equality $h_E=h_M$.

The positive frequency part of the 4D Fourier transform of (\ref{3a}) is 
\begin{equation}
\label{3b}
\mathbf{M}(k)=-i8\pi^2\theta(k^0)\delta(k^2)F(k^0)\mathbf{M}_0,
\end{equation}
where $F(k^0)$ is the 1D Fourier transform of function $f(s)$ in (\ref{3a}). The Fourier transform (see (\ref{1d1})) of the vector potential will then be $\mathbf{A}(k)=\mathbf{k}\times\mathbf{M}(k)$. From (\ref{1e}) and (\ref{3b}) it follows that the current should be defined as $\mathbf{j}=\mathbf{j}_{\perp}=F(k^0)(\mathbf{k}\times\mathbf{M}_0)$ being already orthogonal to $\mathbf{k}$. Substituting this current to (\ref{1j}) produces zero, as is expected because the pulse (\ref{3a}) is linearly polarized.

In order to define a non-linearly polarized pulse the magnetic Hertz vector should be chosen as a weighted sum of two real non-collinear independent vectors $\mathbf{M}^{(1)}$ and $\mathbf{M}^{(2)}$ orthogonal to each other (see, for instance, \cite{fedotov2007exact}), which leads to following expression for the current
\begin{equation}
\mathbf{j}=F^{(1)}(k^0)\left(\mathbf{k}\times\mathbf{M}^{(1)}_0\right)+F^{(2)}(k^0)\left(\mathbf{k}\times\mathbf{M}^{(2)}_0\right),\label{3c}
\end{equation}
where $F^{(1)}(k^0)$ and $F^{(2)}(k^0)$ are two significantly different function. Then the magnetic helicity can be calculated using (\ref{1j}) and (\ref{3c}) and the result is
\begin{equation}
h_M=-\frac{1}{2\pi^2}\int\mathrm{Im}\left[F^{(1)}(k^0)F^{(2)*}(k^0)\right]\mathbf{n}\left(\mathbf{M}_0^{(1)}\times\mathbf{M}_0^{(2)}\right)k^0\,d^3k,\label{3d}
\end{equation}
where $\mathbf{n}=\mathbf{k}/k^0$. Quite obviously the expression (\ref{3d}) is zero meaning that even an polarized arbitrary dipole pulse does not carry any helicity. 

It can be noted that finite electromagnetic pulses of a higher multiplicity (such as quadrupole, octupole, etc.) also carry zero helicity as the vector $\mathbf{n}$ will appear only in odd combinations in (\ref{1j}). On the other hand, the total helicity of a free field, which is a linear combination of pulses of different multiplicities, will generally be different from zero.

Finally, it is useful to calculate the total spin momentum $\mathbf{S}$ and number of photons $N$ for the dipole pulse considered above. This can be done using formulas (17) and (26) from \cite{feshchenko2019number}, which lead to the following result
\begin{eqnarray}
&\mathbf{S}=-\frac{2}{3\pi}\left(\mathbf{M}_0^{(1)}\times\mathbf{M}_0^{(2)}\right)\int\limits_0^{+\infty}\mathrm{Im}\left[F^{(1)}(k^0)F^{(2)*}(k^0)\right]\left(k^0\right)^3\,dk^0,\label{3e1}\\
&N=\frac{2}{3\pi\hbar}\int\limits_0^{+\infty}\left\{\left|F^{(1)}(k^0)\right|^2\left|\mathbf{M}_0^{(1)}\right|^2+2\mathrm{Re}\left[F^{(1)}(k^0)F^{(2)*}(k^0)\right]\mathbf{M}_0^{(1)}\mathbf{M}_0^{(2)}+\left|F^{(2)}(k^0)\right|^2\left|\mathbf{M}_0^{(2)}\right|^2\right\}\left(k^0\right)^3\,dk^0.\label{3e2}
\end{eqnarray}
The spin momentum of a dipole pulse (and of a pulse of any multiplicity) is generally different from zero. This may seem odd, taking into account that the helicity is zero, but can be readily explained by representation (\ref{2j1}): while the spins of all emitted photons can be aligned in the same direction their helicities will have different signs depending on the direction of their wave vectors. 

Let us note that the total helicity of any dipole pulse will remain zero in an arbitrary moving reference frame as well due to the definition (\ref{1a}) and the conservation of $h^i$ in radiative fields.  

\section{Conclusion}
\pst
In this paper a relativistically invariant expression for the total magnetic helicity of the radiative electromagnetic field (e.g. the filed of a laser pulse) is obtained. It does not depend of the choice of the reference frame or gauge and is equal to the total electric helicity of the same field. Both quantities are separately conserved and from this fact it follows that in radiative fields the electric and magnetic vectors are always orthogonal to each other.

The quantization of the electromagnetic field leads to an expression of the radiative field helicity through the field annihilation and creation operators. This reveals the true meaning of the helicity as the algebraic sum of the projection of the spins of individual photons onto their momentums. It is then demonstrated that the total helicity of any dipole pulse is zero whereas its total spin momentum is generally not.

In conclusion, let us note that the concepts of quantum and spin are thought to be absent from the classical physics. Nevertheless, the classical electromagnetic field has invariants that correspond to the operators of the number of photons and the spin angular momentum known in the quantum electrodynamics  \cite{feshchenko2019number}. To these two invariants we in this paper added the third invariant -- the total helicity of the electromagnetic field -- and demonstrated that it similarly has a quantum counterpart -- the helicity operator of the electromagnetic field.

\section*{Acknowledgments}
\pst
The authors are grateful to Dr. I.A. Artyukov for fruitful discussions.

\end{document}